\documentstyle[aps,multicol,epsfig,prb]{revtex}
\input epsf
\def\prb{Phys.\ Rev.\ {\bf B}}
\def\prl{Phys.\ Rev.\ Lett.\/}

\newcommand{\be}{\begin{equation}}
\newcommand{\ee}{\end{equation}}

\newcommand{\e}[1]{\epsilon_{\vec{k}}}

\newcommand{\ba}{\begin{eqnarray}}
\newcommand{\ea}{\end{eqnarray}}

\begin{document}
\title
{Josephson tunnelling spectroscopy of negative $U$
centers}

\author{Vadim Oganesyan$^1$, Steven Kivelson$^1$, Theodore Geballe$^2$, Boris 
Moyzhes$^2$}
\address
{1)  Dept. of Physics,
University of California,
Los Angeles, CA  90095;  2)  Laboratory for Advanced Materials and Dept. of 
Applied Physics, Stanford University, Stanford, CA 94305-4045}
\date{\today}

\maketitle
\begin{abstract}
\
We consider a superconductor-insulator-superconductor (SIS)
junction in which the tunnelling through the insulating barrier
is dominated by a localized ``negative $U$''
center.
We show that the $I_{c}R$ product of the junction depends
sensitively on the spectrum of impurity states, and
in near resonant condiditions
exhibits an anomalously large I$_c$R 
product which can exceed the famous Ambegoakar-Baratoff limit by an 
arbitrarily large factor. The analysis is extended to problems in which
there is an array of negative $U$ centers in the junction. We also discuss
 general reasons to expect significant violations of
the optical conductivity sum rule in most SIS junctions and of 
the Ambegoakar-Baratoff result when the superconductors emerge from a non-Fermi liquid normal state.

\
\end{abstract}
\begin{multicols}{2}

\narrowtext

It has long been known that there exist localized defect
states in solids that can be characterized as ``negative $U$
centers.''  What this means is that the
effective interactions are such that,
\be
U\equiv E(2)+E(0)-2E(1) < 0
\ee
where $E(n)$ is the energy of the state when occupied by $n$
electrons,  {\it i.e.} one can view these impurities as ``pair 
binding'' regions.
The existence of such centers has been found to
explain many of the anomalous properties of various materials in which an impurity or defect
with an odd number of electrons produces a diamagnetic (rather than the naively expected
paramagnetic) response.  In particular, this model was discussed by Anderson\cite{pwa}
in connection with dangling bonds in chalcogonide glasses.  

An obvious question
arises whether the effective attraction between electrons implied by Eq. (1) can be, in
some way, harnessed to provide a mechanism for pairing in a superconductor.  The trouble,
of course, is that the negative $U$ centers with which we are familiar are
highly localized, and the materials in which they occur 
are good insulators rather than superconductors.
However, in a  multicomponent
system, it is possible\cite{ekz} for electron itineracy to derive
from one component which is proximity coupled to a set of localized  negative
$U$ centers, from which the pairing arises.  Indeed, superconductivity
with
$T_c=1.5$K has been observed\cite{pbte} in semiconducting PbTe doped with
around 1\% Tl, a well known\cite{negU} negative $U$ center.  Along these
lines, two of us\cite{geballe} have recently proposed that negative
$U$ centers between the copper-oxide planes may play a role in enhancing $T_c$ in certain
cuprate high temperature superconductors, most notably in the Hg based materials.

In the present paper, we address the less ambitious, but
related issues:  What is the effect of {\em weak} coupling
between a conventional superconductor and a negative $U$ center?  Can 
Josephson tunnelling be used as a spectroscopic probe to {\em detect}
the presence of negative $U$ centers?  Manifestly, in the weak
coupling limit, the presence of a negative $U$ center
can have little or no effect
on the strength of the superconducting state, itself. However, as we 
will show 
below, the incipient pairing on the negative $U$ center leads to an 
anomalous enhancement of the I$_c$R product;  it can
exceed the famous Ambegoakar-Baratoff limit\cite{ab} by an arbitrarily large 
factor. 
By sweeping the chemical potential on the impurity site (perhaps
by applying a suitable gate voltage) the impurity level can be moved
closer and further from resonance, which leads to predictable
spectrocopic variations of the critical current.  We also discuss the
generalization of these results and their possible pertinence to 
recent experiments in layered superconductors.

{\bf The Case of a single negative U center:}
To begin, we consider a tunnel junction between two
pieces of bulk superconductor through a barrier containing a
single negative U center.  Specifically, we consider the
model Hamiltonian
\be
H=H_L + H_R + H_U + H_{tun}
\ee
where $H_L$ and $H_R$ are the Hamiltonians of the left and
right superconductors, which we take to be identical
superconductors and well approximated by the
Bougoliubov-deGennes equations\cite{schrieffer}, {\it i.e.}
\be
H_{L}=\sum_{k\sigma}(\epsilon_{k}-\mu)L_{k\sigma}^{\dagger}L_{k\sigma}
+\Delta (L_{k\sigma}^{\dagger}L_{-k-\sigma}^{\dagger}+h.c.).
\ee
and the negative U center is described by 
\be
H_U= -|U|
c_{\uparrow}^{\dagger}c_{\uparrow}c_{\downarrow}^{\dagger}
c_{\downarrow}
+
(\varepsilon-\mu)\sum_{\sigma} c_{\sigma}^{\dagger}c_{\sigma}.
\ee
Here, $c_{\sigma}^{\dagger}$ creates an electron on
the impurity site with spin polarization $\sigma$
and
$R_{\sigma}^{\dagger}$ and $L_{\sigma}^{\dagger}$
create, respectively, an electron at the edge of the
junction in the right and left superconductors.  
The tunneling Hamiltonian (which we treat as a small
perturbation) is
\be
H_{tun}= -\sum_{\sigma} [t_{L}L_{\sigma}^{\dagger}c_{\sigma}
+
t_{R}R_{\sigma}^{\dagger}c_{\sigma}+ h.c.].
\ee
(Of course, other interactions may be important under some 
circumstances, including direct tunnelling through the junction
(not involving the impurity) and more complicated pair-tunelling and
other interaction terms;  under appropriate circumstances, the present
model includes the most important interactions, and it is sufficient
for our present purposes.)

In the absence of the tunnelling term
the eigenstates of the impurity are singly occupied, unoccupied or 
doubly occupied (in a spin singlet). The ground state is {\it always} 
one of the latter two, so it is 
more convenient to
express the results in terms of the single and two-particle 
excitation energies
$\epsilon_1$ and
$\epsilon_2$, both positive,  rather than to work directly with $U$
and
$\varepsilon$. For example,  for $\varepsilon-\mu>0$ and
$|U|<2(\varepsilon-\mu)$, the groundstate  has no particles in it
and $\epsilon_1=\varepsilon-\mu$ and 
$\epsilon_2=2(\varepsilon-\mu)-|U|$.
Although all of our results can be derived for the entire range of 
parameters 
we are specifically interested in the case where $\epsilon_1$ is 
large. In the single impurity case we will use 
$\epsilon_1/\Delta\gg1$ to simplify the expressions.

It is now a straightforward exercise to compute the
zero temperature
Josephson coupling to lowest (4${th}$) order in powers of
$H_{tun}$:
\be
J=
 \Delta \left[\frac{\pi N_F t_Lt_R} {\epsilon_1}\right]^2
\left[1+
\frac{\Delta}{\epsilon_2}
\frac{\log^2(2\epsilon_1/\Delta)}{\pi^2}
\right],
\label{t4}
\ee
where $N_F$ is the normal state 
density of states at the Fermi energy.
 The sign\cite{sign} of the
$J$ is such as to favor alignment of the phases ($\theta_{L}$ and 
$\theta_R$) across the
junction;  {\i.e.} the Josephson energy is
$E_{J}=-J\cos(\theta_L-\theta_R)$.
  As usual, the critical current is simply
$I_c=2eJ/c\hbar$.

The Josephson coupling is the sum of two terms:  The
first term is the contribution to the pair tunnelling across
the junction from processes in which  first one electron and
then the other passes from one side  of the junction to the
other.  The second term comes from processes in which the
second electron enters the junction before the first has
left it.  In this sense, one can think of the first term as
being the textbook\cite{schrieffer} result with an effective
tunnelling interaction $T_{LR}= t_Lt_R/\epsilon_1$.  However,
the second term, which involves correlated pair tunnelling,
is a qualitatively new effect.
Being proportional to $\frac{\Delta}{\epsilon_2}$, it
diverges at the point at which the unoccupied and doubly
occupied impurity states are  degenerate (the impurity becomes 
partially occupied), {\it i.e.} a resonance
condition for pairhopping.
Note that, because the junction conductance is
proportional to the square of the amplitude for single-particle transmission
across the barrier, there is no analagous correlated pair-tunnelling term which
contributes to $1/R$.

By itself, this
expression is not terribly illuminating.  However, as usual,
we can express $J$ in units of the normal-state junction
conductance, $1/R$, so that the highly detail dependent
factors of the tunnelling matrix elements cancel, and we are
left with a quantity, with units of energy, which in some
sense measures how big the Josephson coupling is in natural
units.  $R$ can be computed using a straightforward
generalization of standard perturbative methods. It is independent of 
how far the impurity is from resonance ($\epsilon_{2}$). In the limit
of small
$\frac{\Delta}{\epsilon_1}$, the result, the principal result of this 
paper,
is
\be
eI_c R = \frac {\pi \Delta} {2} \left[ 1 +\frac {\Delta} 
{\pi^2\epsilon_2}
\log^2(2\epsilon_1/\Delta)\right]
\label{icr}
\ee

Note that far from resonance, when the second term is absent, 
this result reproduces the standard
Ambegoakar-Baratoff\cite{ab} relation.  
However,  near the resonance, 
where the unoccupied and doubly occupied states
of the impurity are nearly degenerate with each other, the
negative $U$ center gives rise to a dramatically enhanced
$I_cR$ product.
When the resonance condition is
too nearly satified, {\it i.e.} $\epsilon_2/\Delta < N_F t^2/\epsilon_1$,
the perturbative expression is no longer useful.

Luckily, this single impurity problem can be solved for 
arbitrary $\epsilon_2$ (but still perturbatively in tunnelling): 
integrating (projecting) out the singly occupied impurity states and 
the two superconductors reduces the problem to a rather simple 
effective Hamiltonian
\be
H^{eff}=\left({0 \atop \delta}\ {\delta^* \atop \epsilon_2} \right)+E_J^{(1)}
\ee
for the unoccupied and doubly occupied states in the effective 
proximity field 
\be
\delta=\frac{\Delta N_F \log(2\epsilon_1/\Delta)}{\epsilon_1}(t^2_L 
e^{i\theta_L}+
t^2_R e^{i\theta_R})
\ee
and $E_J^{(1)}$ is the contribution from the first term in Eq.  \ref{t4} to the
Josephson energy.
The ground state energy
\be
E_{J}(\theta_L-\theta_R)=\frac{\epsilon_2}{2}-\sqrt{\left(\frac{\epsilon_2}{2}\right)^2+\delta^*\delta}
+E_J^{(1)}
\ee
 can now be used not only to obtain the altogether larger, ${\cal 
O}(t^2)$, contribution right at the resonance ($\epsilon_2 = 0$) 
\be
E_{J}^{res}=\frac{\Delta N_F 
\log(2\epsilon_1/\Delta)}{\epsilon_1}\sqrt{t^4_L+t^4_R+2(t_Lt_R)^2\cos(\theta_L-\theta_R)},
\ee
 but also to smoothly connect with Eq. \ref{t4}.

In principle, when this kind of an effectively single impurity 
junction is realized, the characteristic energies, 
$\epsilon_2$ and $\epsilon_1$, can be varied by applying a voltage in 
the
junction, so that the resonant condition can be tuned.
In many circumstances one is faced with a 
collection of such impurities and so it is worth generalizing the 
derivation for an extended system.

{\bf Tunnelling through a correlated region:}
The general structure of the perturbative calculation is 
unchanged. For the single impurity the result is determined 
 by purely local correlations (both of the superconductors and, 
trivially, the impurity) and therefore a single parameter, 
$\frac{\Delta}{\epsilon_{2}}$, controls the physics.
For an extended barrier, a far more 
diverse roster of 
possibilities exists due to the interplay of disorder and coherence 
scales of the superconductors and in the barrier. 

In the tunnelling 
Hamiltonian we now allow for a non-uniform (though still 
short-ranged) hopping:
\be
H_{tun}=\sum_{\sigma}\int d \vec{r} 
[t_{L}(\vec{r})L_{\sigma}^{\dagger}(\vec{r})c_{\sigma}(\vec{r})
+
t_{R}(\vec{r})R_{\sigma}^{\dagger}(\vec{r})c_{\sigma}(\vec{r})+ h.c.].
\ee
The superconductors are still described via Bogoliubov-deGennes 
Hamiltonian or, more generally, by their anomalous (single particle) 
green functions ${\cal F}_{L/R}(\vec{r},\tau)$\cite{schrieffer}.  The impurity region is
described entirely by its (time-ordered) two particle 
correlation function $\langle  T_{\tau} c_\downarrow(1) c_\uparrow(2) 
c^\dagger_\uparrow(3) c^\dagger_\downarrow(4) \rangle$, 
where $1\ldots4$ is a short-hand for space and time coordinates, $(\vec
r_1,\tau_1)$.

We now make a crucial simplifying assumption in describing the 
nonsuperconducting layer. 
We will take its single electron green function to be
 short ranged in space and time. 
This is certainly well justified when there is a gap or, more 
generally, whenever the relevant electron energy scales
are large compared to $\Delta$.   

More specifically, we can always define
the pair ($P_2$) and partidle-particle propagators ($P_1$ and $P_0$) by decomposing
the two particle correlation function as
\end{multicols}
\widetext
\noindent
\setlength{\unitlength}{1in}
\begin{picture}(3.375,0)
  \put(0,0){\line(1,0){3.375}}
  \put(3.375,0){\line(0,1){0.08}}
\end{picture}
\be
\langle  T_{\tau} c_\downarrow(1) c_\uparrow(2) 
c^\dagger_\uparrow(3) c^\dagger_\downarrow(4) \rangle =
G(1,2)G(3,4)P_2(1,2;3,4) + G(1,4)G(2,3)P_1(1,4;2,3) + G(1,3)G(2,4)P_0(1,3;2,4)
\label{Palpha}
\ee
\hfill
\begin{picture}(3.375,0)
  \put(0,0){\line(1,0){3.375}}
  \put(0,0){\line(0,-1){0.08}}
\end{picture}
\begin{multicols}{2}
\noindent{where} $G (\vec{r}_1,\vec{r}_2,\tau_1-\tau_2)=\langle T_{\tau}
c^\dagger_\uparrow(\vec{r}\,\tau)c_\uparrow(\vec{r}\ 
',\tau')\rangle$ is the
imaginary 
time-ordered
Green function, and where $P_1\to 0$, $P_3\to 0$, and $P_0\to 1$ at large
distances.  The pertrubative expression for the Josephson coupling
involves integrals over this quanatity.   We will simplify this expression
by making two physically motivated assumptions. Firstly, we will assume
that the integral of  
$G$  and any other less strongly peaked function $f$ 
can be simplified as follows
\be
\int_{0}^{\beta}d \tau'\int d\vec{r}\ ' G(\vec{r},\vec{r}\ 
',\tau')
f(\vec{r}\ ',\tau') \rightarrow
\frac{1}{\epsilon}f(\vec{r}, \frac{1}{\epsilon}),
\label{epsilon}
\ee
where $\epsilon$ is a characteristic single particle excitation 
energy (analogous to $\epsilon_{1}$ above).  
If $f(\vec r,0)$ is finite, we can let $f(\vec
r,1/\epsilon)\rightarrow f(\vec r,0)$, but if $f$ is 
weakly singular as $\tau \to 0$,
$1/\epsilon$ plays the role of a cutoff - see below.  
Among other things, this permits us to substitute $P_{\alpha}(1,2;3,4)
\to P_{\alpha}(1,1;4,4)\equiv {\cal P}_{\alpha}(1,4)$ in Eq. \ref{Palpha}.
Secondly, since we are interested in a barrier region which supports
significant pairing fluctuations, we will assume that all the interesting
correlation effects are reflected in the behavior of $P_2$, an will
consequently make the simplifying approximations $P_0=1$ (no significant
charge density wave fluctuations) and
$P_1=0$ (no significant magnetic fluctuations). 

With these simplifying approximations, 
we can bring the analog of Eq. \ref{t4},
$J=J_1+J_2$, into a  somewhat more manageable form with 
\be
J_1={2} \int d\vec{r}d\vec{r}\ ' 
 T_{LR}(\vec{r})
T_{LR}(\vec{r}\ ') \int_0^\beta d\tau 
|{\cal F}(\vec{r}-\vec{r}\ ',\tau)|^2
\ee
\be
J_2=
\frac{2|{\cal F}(\vec{0},\frac{1}{\epsilon})|^2}{\epsilon^2}\int 
d\vec{r}d\vec{r}\ '
 t^2_L(\vec{r})
t^2_R(\vec{r}\ ')\int_0^\beta d\tau {\cal P}_1(\vec{r},\vec{r}\ ',\tau),
\ee
\noindent
where $T_{LR}(\vec{r})\equiv t_L(\vec{r})t_R(\vec{r})/\epsilon$.
 Notice that the first term, as before, only depends on the 
parameters of the tunneling region through the typical single 
particle energy, $\epsilon$, and, so can be viewed as a direct 
tunnelling term but with an effective hopping matrix, $T_{LR}(\vec r)$. More 
importantly, the hopping elements, $t_L(\vec{r})$ and $t_R(\vec{r})$, enter  $J_1$ 
and $J_2$ very differently, and so disorder will generally have a very different 
effect on the two contributions. 

To make our analysis concrete, we adopt a simple model
in which there are random variations of $t_L(\vec r)$ and $t_R(\vec r)$, but
no correlations between $t_L$ and $t_R$:  
\be
\overline{t_L(\vec r)t_L(\vec r')}=
t_L^2[\alpha_L + (1-\alpha_L)\exp(-|\vec r-\vec r'|/\xi_L)]
\ee
(and similarly for $t_R$),
but $\overline{t_L(\vec r)t_R(\vec r')}=0$,
where $\overline{t_L(\vec r)}=t_L\sqrt{\alpha_L}$ signifies the configuration average.
For simplicity, we ignore all other sources of
disorder.

In terms of this model, it is possible to obtain expressions for the various
contributions to the Josephson coupling in terms of various averages over the
various propagators.  These expressions are simple in terms of the 
implicitly  defined  intrinsic coherence lengths
\ba
\int d \tau d^{d-1}r {\cal P}_1 (\vec r,0,\tau)
&=&\frac{\xi_2^{d-1}}{\epsilon_2},\\
\label{epsilonxi2}
\int d \tau d^{d-1}r  |{\cal F} (\vec r,0,\tau)|^{2}
&=&\frac{\xi_0^{d-1}}{\Delta}[\Delta N_F]^2,
\ea 
where, in addition
\be
\int d\tau  {\cal F}(\vec 0,\tau)  = \frac {\pi \Delta N_F} {\Delta}, \ \ {\rm and} \ \
{\cal F}(\vec 0,\frac 1 {\epsilon}) =\Delta N_F \log[\epsilon/\Delta]. 
\label{calF}
\ee
It is interesting to remark that the various expressions involving ${\cal F}$, which we
have evaluated here in the context of BCS mean-field theory, depend on the 
behavior of the superconducting leads at distances {\em short} compared to
$\xi_0$.  Here, as we discuss below, the correlation functions intimately
reflect the fact that the {\em normal} state of the leads is a Fermi liquid.

We do not display the complete result, here, since it is long.  In the limit of weak disorder,
$\alpha\to 1$, the configuration averaged Josephson energy per unit area (area $\equiv
L^{d-1}$) is readily evaluated
\ba
{\cal J}&=&
\frac{2   \Delta( \pi N_F t_L t_R)^2}{\epsilon^2} \left(
{\xi_0^{d-1}}+
\frac{\xi_2^{d-1}\Delta }{\pi^2 \epsilon_2}\log^2({\epsilon/\Delta})
\right).
\label{nodisorder}
\ea
Here, the ratio of the normal contribution and
correlated pair tunnelling contributionsv to the Josephson coupling is proportional to
$(\epsilon_2/\Delta)(\xi_0/\xi_2)^{d-1}$;  unless the pair
correlations are at quite low energy and extend over quite long distances, the anomalous (${J}_2$) pair-tunnelling
term will be insignificant.

Now we consider a large disorder limit, $\alpha\to 0$,
and for simplicity assume as well that
$\xi_L$, $\xi_R\ll\xi_2$,
$\xi_0$. In this case,
\ba
{\cal J}=
\frac{2 c_L c_R  \Delta( \pi N_F t_L t_R)^2}{\epsilon^2} \Bigg[&&
\Upsilon_1\left(\frac {\xi_L\xi_R}{\xi_L+\xi_R}\right)^{d-1} \\
\label{disorder}
&&+\Upsilon_2\frac{\xi_2^{d-1}\Delta }{\pi^2 \epsilon_2}\log^2({\epsilon/\Delta})
\Bigg],\nonumber
\ea
where $\Upsilon_J$ are geometric factors of order 1.  Here, so long as $\xi_2 \gg \xi_L$,
$\xi_R$, there is a large enhancement of the correlated pair tunnelling term.  The physics of
this enhancement is very simple - if $\xi_2$ is large, a pair can tunnel into the barrier where
$t_R$ is large, then propagate along the barrier, and finally tunnel out where $t_L$ is
large.  This spatial structure amplifies the sort of resonance effects we found in the single
impurity case.

{\bf Discussion:}
With regards to application of these ideas to layered superconductors,
our conclusions are intriguing, but not unambiguous. 
 Large I$_c$R 
products have recently been measured along the c-axis in various 
layered 
superconductors\cite{basov}, as have dramatic shifts of optical 
spectral weight over large ranges of energies. 
The two phenomena are thought  by many to be related.
We find that all simple models of SIS junctions (and more 
generally\cite{marel}) exhibit
large spectral weight shifts leading to apparent 
sum rule violations as the system enters the superconducting state. 

To see this, consider the case of a
simple junction,{\it i.e.} $H_{tun}= -\sum_{\sigma} [t L_{\sigma}^{\dagger}R_{\sigma}
+ h.c.]$.  Rather than computing the actual optical
conductivity 
we use the single-band
version of the optical sum rule (which relates the
integrated spectral weight in a given band to the
expectation value of the kinetic energy) to compute the
integrated oscillator strength for the normal junction and
the superconducting junction.  The result is that
\be
\int d\omega [\sigma_N(\omega)-\sigma_S(\omega)]
=4 |t|^2 N_F^2 \Delta \log \frac{W}{\Delta}.
\ee
where $\sigma_N$ and $\sigma_S$ are, respectively, the
frequency dependent junction conductance with superconducting
and non-superconducting leads.
Since the f-sum rule must ultimately be satisfied,
presumably the correct interpretation of this result is that
spectral weight is transferred over large energy ranges,
beyond those described by our model, {\it i.e.} on energy
scales on the order of the band-width, which is obviously
much larger than $\Delta$.  Since this occurs even in the
absence of an $I_cR$ anomaly, we are at present unable to
make a clear statement concerning the expected spectral
consequences of such an anomaly.
Thus, the observed apparent sum-rule
violations cannot be definitively linked with 
the presence or absence of interesting
correlations in the barrier region.
And, unfortunately, the experimental data on large I$_{c}$R products
in the high temperature superconductors, while very striking indeed,
may potentially be hard to interpret as well, as it is intertwined 
with
pseudo-gap phenomena which onset well above the superconducting 
transition temperature.

Perhaps yet a potentially more vexing concern has to do with the non-universality 
of the I$_c$R product even when tunnelling through an uncorrelated 
insulator.  
The problem is that the result is strongly influenced by the  short 
distance/time properties of the superconductor. In weak coupling 
these are inherited from the ``normal'' state. We have already seen one 
manifestation of this: the logarithm in the anomalous term is 
inherited from the underlying Fermi Liquid (this is the same log 
responsible for the superconducting instability). More generally, 
even the simple proportionality, ${\cal F}\sim \Delta N_F$, very 
familiar from the BCS theory (and in part responsible for the 
Ambegoakar/Baratoff result), between the order parameter (${\cal F}$) 
and the mean field ($\Delta$) need not be generic\cite{lan} either deep enough 
in the superconducting phase or when the normal phase is not Fermi 
Liquid like.

In summary, we have considered a problem of tunnelling through a 
single negative U impurity and an extended region with incipient 
pairing.
We find,  under favorable
circumstances, anomalous enhancement of the Josephson 
coupling due to correlated pair tunnelling processes.
 
{\bf Aknowledgements:} This work was supported, in part, by NSF grant
\#DMR-0110329 at UCLA (VO and SAK).  The work at Stanford was 
supported in part by the Airforce Office of Scientific Research.  SAK and VO
wish to acknowledge the hospitality of the Laboratory for Advance Materials
(now, the Geballe Laboratory for Advanced Materials) at Stanford University
where this work was initiated.

\end{multicols}
\end{document}